\documentclass{article}

\input pz.sty
\input epsf.sty
\usepackage{longtable}

\begin{document}

\PZhead{3}{32}{2012}{10 July}

\PZtitletl{Photometric observations and preliminary modeling}{of 
type IIb supernova 2011\lowercase{dh}}

\PZauth{D.Yu. Tsvetkov$^1$, I.M. Volkov$^{1,2}$, E.I. Sorokina$^1$,
S.I. Blinnikov$^{3,1}$, N.N. Pavlyuk$^1$, G.V. Borisov$^4$} 
\PZinst{Sternberg Astronomical Institute, Lomonosov Moscow 
State University, University Ave. 13,
119992 Moscow, Russia}
\PZinst{Astronomical Institute of the Slovak Academy of Sciences, 059 60 
Tatranska Lomnica, Slovak Republic}
\PZinst{Institute for Theoretical and Experimental Physics, 
Bol'shaya Cheryomushkinskaya Str. 25,
117259 Moscow, 
Russia}
\PZinst{Crimean Laboratory of Sternberg Astronomical Institute, Lomonosov
Moscow State University, Nauchnyi,
Crimea, Ukraine}

\begin{abstract}

CCD {\it UBVRI} photometry is presented for type IIb SN 2011dh for about
300 days. 
The main photometric parameters are derived and the 
comparison with SNe of similar types is reported. 
The light curves are similar to those for SN IIb 2008ax, but the
initial flash is stronger and very short, and there are humps on
the light curves in $U$ and $B$ at the onset of linear decline.
Preliminary modeling 
is carried out, and the results are compared to the quasi-bolometric
light curve and to the light 
curves in {\it UBVRI} bands.  

\end{abstract}

\medskip
\medskip
\bigskip
\PZsubtitle{\bf Introduction}
On May 31, 2011 a supernova exploded in bright nearby spiral galaxy 
M51 (NGC5194, Whirlpool galaxy).
The outburst, designated SN 2011dh, was promptly
discovered independently by several amateurs and the Palomar
Transient Factory (see CBET No.2736 and Arcavi et al. (2011b) for
details).
The early discovery of such a nearby SN facilitated numerous 
follow-up studies. Early spectra and light curve indicated 
SN 2011dh to belong to the class of stripped-envelope
core-collapse SN, designated as type IIb (Arcavi et al., 2011a, 2011b).
The progenitor or progenitor system was identified in archival 
images obtained by the HST (Li and Filippenko, 2011), although its nature
remains controversial. Maund et al. (2011) suggest it was
a yellow supergiant with initial mass about 13$M_{\odot}$, while 
Van Dyk et al. (2011) prefer higher mass in the range 18-21 $M_{\odot}$.
The variability of the candidate progenitor was reported
by Szczygiel et al. (2012).
Multi-wavelength follow-up observations in the radio, millimiter, X-ray and
gamma-ray bands suggest a compact progenitor with $R\approx 10^{11}$ cm,
which is inconsistent with the radius of the yellow supergiant, so 
this star may be a binary companion of presupernova
or even unrelated to the SN (Soderberg et al., 2012). 
Radio observations were reported also by Krauss et al. (2012),
Marti-Vidal et al. (2011) and Bietenholz et al. (2012).
Vinko et al. (2012) presented optical spectroscopy and photometry of
SN 2011dh and applied the EPM method to derive the distance of 
8.4 Mpc for M51.

\newpage
\PZsubtitle{\bf Observations and reductions}
On June 1 a sequence of unfiltered images of M51 was obtained with 
the 192-mm telescope (hereafter C19), equipped with FLI PL16803 CCD camera,
at Crimean Observatory of Sternberg Astronomical Institute (SAI).
53 frames were obtained in the period 19:25--20:31 UT.
5 days later we started regular monitoring of SN 2011dh and
continued observations until 2012 April 4. CCD images in 
Johnson-Cousins {\it UBVRI} bands were obtained with the
following instruments: the 15-cm and 60-cm telescopes of Astronomical Institute
of Slovak Academy of Sciences at Tatranska Lomnica (S15, S60),
equipped, respectively, with SBIG ST-10XME and Princeton Instruments
VersArray F512 CCD cameras;
the 60-cm reflector of Crimean Observatory of SAI
(C60) with Apogee AP-47p camera; the 70-cm reflector of SAI
in Moscow (M70) with Apogee AP-7p camera; 60-cm reflector of 
Simeiz
Observatory (K60) with VersArray F512 camera, and 1-m reflector of 
Simeiz Observatory, equipped with either VersArray F512, or 
VersArray B1300 cameras (K100a, K100b). 

The standard image reductions and photometry were made using IRAF.\PZfm 
\PZfoot{IRAF is distributed by the National Optical Astronomy Observatory,
which is operated by AURA under cooperative agreement with the
National Science Foundation}

The galaxy background around SN 2011dh is quite smooth,
nevertheless we applied image subtraction for all of the frames 
obtained later then 2011 September 20.
The template images were constructed from frames obtained at C60
while carrying out observations of SN 2005cs.

The magnitudes of the SN
were derived by PSF-fitting relative to 
a sequence of local standard stars. 
The image of SN 2011dh and comparison stars is shown in Fig. 1. 
\PZfig{11cm}{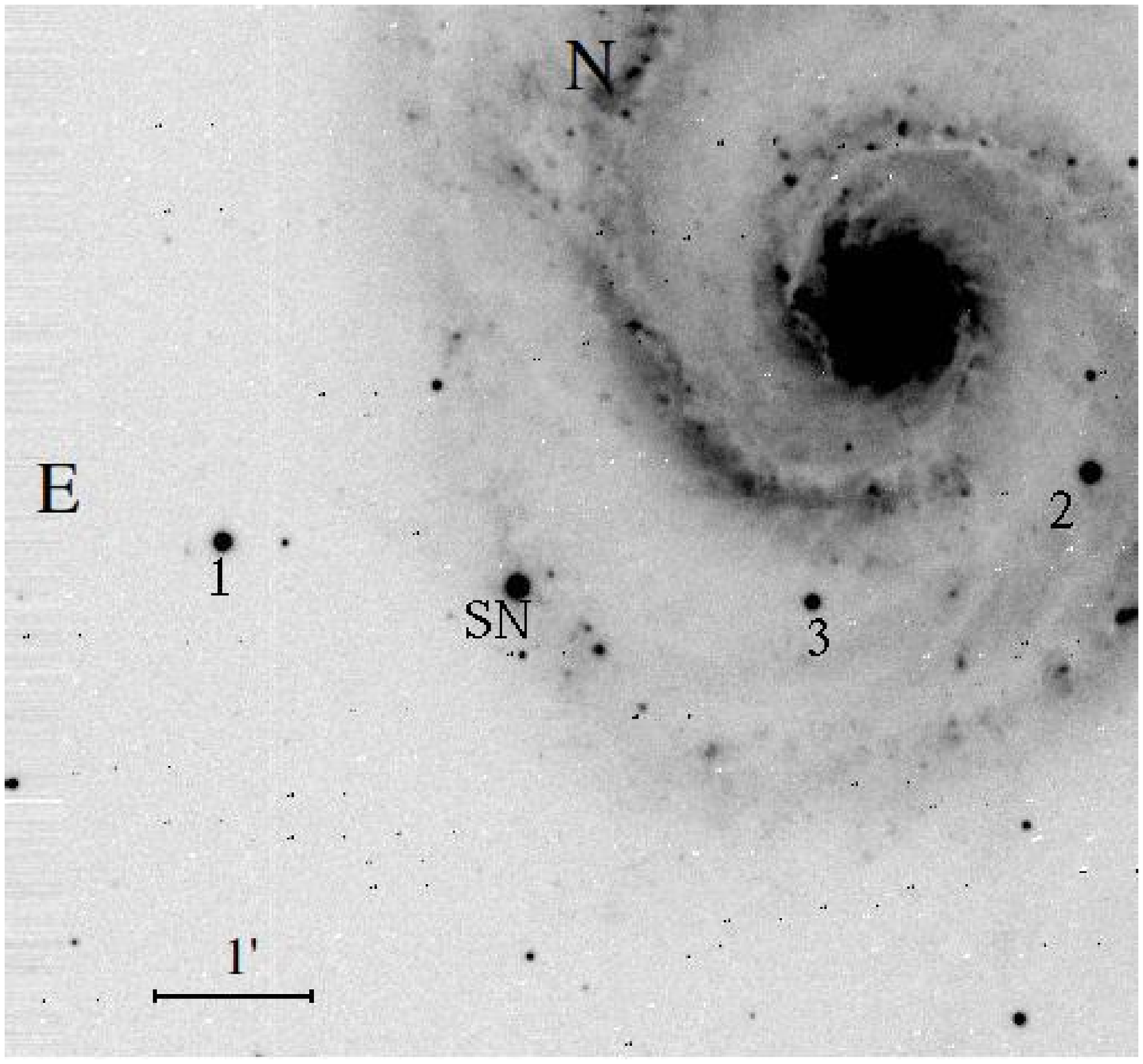}{SN 2011dh with local standard
stars}
The magnitudes of these stars were taken from Pastorello et al. (2009)
On the images with larger field of view we also used more 
distant comparison stars, also from Pastorello et al. (2009).

The results of observations of the SN are presented in Table 1.
We did not detect significant variations of brightness on the
sequence of images obtained on June 1, so the averaged values, 
calibrated by $V$ magnitudes, are reported in Table 1.

\begin{center}
\begin{longtable}{cccccccccccl}
\caption{Observations of SN 2011dh}\\
\hline
JD 2450000+ & $U$ & $\sigma_U$ & 
$B$ & $\sigma_B$ & $V$ & $\sigma_V$ & $R$ & $\sigma_R$ &
$I$ & $\sigma_I$ & Tel.\\
\hline
\endfirsthead
\multicolumn{12}{c}{\tablename\ \thetable{} -- continued from previous page}\\
\hline
JD 2450000+ & $U$ & $\sigma_U$ &
$B$ & $\sigma_B$ & $V$ & $\sigma_V$ & $R$ & $\sigma_R$ &
$I$ & $\sigma_I$ & Tel.\\
\hline
\endhead

\hline
\endfoot

\hline
\endlastfoot

5714.32&       &      &      &      & 13.54 & 0.03 &       &     &       &      & C19 \\
5714.35&       &      &      &      & 13.52 & 0.03 &       &     &       &      & C19 \\
5719.34& 14.67 & 0.06 &14.54 & 0.02 & 13.94 & 0.02 & 13.64 & 0.01&  13.61&  0.01& M70 \\
5723.37& 13.85 & 0.04 &13.79 & 0.02 & 13.17 & 0.02 & 12.89 & 0.01&       &      & K100a\\ 
5724.37& 13.83 & 0.04 &13.70 & 0.02 & 13.03 & 0.02 & 12.77 & 0.01&       &      & K100a\\ 
5725.31& 13.65 & 0.04 &13.60 & 0.02 & 12.92 & 0.02 & 12.65 & 0.01&  12.56&  0.01& K100a\\ 
5726.31& 13.62 & 0.04 &13.55 & 0.02 & 12.87 & 0.02 & 12.65 & 0.01&  12.49&  0.01& K100a\\ 
5727.28& 13.60 & 0.04 &13.53 & 0.02 & 12.79 & 0.02 & 12.51 & 0.01&  12.40&  0.01& K100a\\ 
5729.31& 13.52 & 0.04 &13.42 & 0.02 & 12.67 & 0.02 & 12.39 & 0.01&  12.26&  0.01& K100a\\ 
5730.31& 13.53 & 0.04 &13.42 & 0.02 & 12.64 & 0.02 & 12.34 & 0.01&  12.22&  0.01& K100a\\ 
5731.30& 13.53 & 0.04 &13.38 & 0.02 & 12.60 & 0.02 & 12.30 & 0.01&  12.17&  0.01& K100a\\ 
5731.32& 13.49 & 0.07 &13.32 & 0.02 & 12.60 & 0.02 & 12.31 & 0.01&  12.17&  0.01& K100a\\ 
5733.31& 13.59 & 0.04 &13.41 & 0.02 & 12.55 & 0.02 & 12.27 & 0.01&  12.11&  0.01& K100a\\ 
5734.29& 13.66 & 0.04 &13.42 & 0.02 & 12.55 & 0.02 & 12.21 & 0.01&  12.07&  0.01& K100a\\ 
5734.37& 13.63 & 0.08 &13.36 & 0.02 & 12.60 & 0.04 & 12.26 & 0.03&  12.09&  0.04& C60\\ 
5735.27& 13.76 & 0.04 &13.48 & 0.02 & 12.57 & 0.02 & 12.22 & 0.01&  12.06&  0.01& K100a\\ 
5735.33& 13.78 & 0.06 &13.43 & 0.02 & 12.60 & 0.02 & 12.25 & 0.03&  12.08&  0.04& C60\\ 
5736.28& 13.90 & 0.04 &13.59 & 0.02 & 12.61 & 0.02 & 12.24 & 0.01&  12.07&  0.01& K100a\\ 
5737.28& 14.16 & 0.04 &13.72 & 0.02 & 12.66 & 0.02 & 12.27 & 0.01&  12.08&  0.01& K100a\\ 
5741.28& 15.15 & 0.05 &14.28 & 0.02 & 13.00 & 0.02 & 12.49 & 0.01&  12.25&  0.01& K100a\\ 
5743.28& 15.44 & 0.04 &14.57 & 0.02 & 13.19 & 0.02 & 12.61 & 0.01&  12.33&  0.01& K100a\\ 
5744.32& 15.71 & 0.05 &14.79 & 0.02 & 13.27 & 0.02 & 12.67 & 0.01&       &      & K100a\\ 
5744.33&       &      &14.63 & 0.05 & 13.35 & 0.02 & 12.72 & 0.02&  12.43&  0.03& C60\\ 
5746.28& 15.95 & 0.04 &14.89 & 0.02 & 13.41 & 0.02 & 12.79 & 0.01&  12.47&  0.01& K100a\\ 
5747.27& 16.06 & 0.06 &14.99 & 0.02 & 13.48 & 0.02 & 12.82 & 0.01&  12.48&  0.01& K100a\\ 
5749.30& 16.36 & 0.12 &15.07 & 0.02 & 13.59 & 0.02 & 12.91 & 0.01&  12.57&  0.01& K100a\\ 
5749.34&       &      &14.99 & 0.02 & 13.65 & 0.02 & 12.95 & 0.02&  12.60&  0.03& C60\\ 
5750.29& 16.37 & 0.11 &15.11 & 0.02 & 13.64 & 0.02 & 12.96 & 0.01&  12.62&  0.01& K100a\\  
5751.26& 16.41 & 0.04 &15.21 & 0.02 & 13.69 & 0.02 & 13.00 & 0.01&  12.63&  0.01& K100a\\ 
5752.35& 16.46 & 0.04 &15.24 & 0.02 & 13.75 & 0.02 & 13.05 & 0.01&  12.67&  0.01& K100a\\ 
5755.35&       &      &15.34 & 0.02 & 13.94 & 0.02 & 13.19 & 0.02&  12.79&  0.03& C60\\ 
5757.34&       &      &15.31 & 0.04 & 14.00 & 0.02 & 13.26 & 0.02&  12.84&  0.02& C60\\ 
5783.28& 16.19 & 0.17 &15.68 & 0.03 & 14.53 & 0.04 & 13.85 & 0.01&  13.41&  0.02& C60\\ 
5786.31&       &      &15.66 & 0.03 & 14.60 & 0.04 & 13.91 & 0.02&  13.48&  0.02& C60\\ 
5788.28&       &      &15.74 & 0.05 & 14.62 & 0.04 & 13.95 & 0.03&  13.53&  0.03& C60\\ 
5789.26&       &      &15.69 & 0.03 & 14.64 & 0.02 & 13.97 & 0.02&  13.55&  0.03& C60\\ 
5790.24&       &      &15.78 & 0.04 & 14.62 & 0.02 & 13.99 & 0.01&  13.56&  0.02& C60\\ 
5808.24&       &      &15.94 & 0.02 & 14.98 & 0.02 & 14.39 & 0.02&  13.99&  0.02& C60\\ 
5811.22&       &      &15.98 & 0.03 & 15.03 & 0.02 & 14.45 & 0.02&  14.07&  0.02& C60\\ 
5817.22&       &      &16.01 & 0.03 & 15.16 & 0.02 & 14.62 & 0.02&  14.07&  0.03& K100b\\ 
5818.22& 16.39 & 0.06 &16.06 & 0.02 & 15.20 & 0.02 & 14.61 & 0.02&  14.09&  0.03& K100b\\ 
5819.21& 16.71 & 0.07 &16.10 & 0.02 & 15.23 & 0.01 & 14.66 & 0.01&  14.10&  0.03& K100b\\ 
5820.21& 16.57 & 0.05 &16.12 & 0.02 & 15.22 & 0.02 & 14.66 & 0.02&  14.11&  0.03& K100b\\ 
5821.21& 16.62 & 0.08 &16.08 & 0.02 & 15.22 & 0.02 & 14.68 & 0.02&  14.13&  0.03& K100b\\ 
5822.21& 16.66 & 0.08 &16.14 & 0.02 & 15.24 & 0.02 & 14.69 & 0.02&  14.17&  0.03& K100b\\ 
5823.24& 16.70 & 0.09 &16.15 & 0.02 & 15.27 & 0.02 & 14.73 & 0.03&  14.16&  0.02& K100b\\ 
5825.20& 16.63 & 0.07 &16.17 & 0.03 & 15.34 & 0.02 & 14.75 & 0.02&  14.25&  0.03& K100b\\ 
5830.21& 16.82 & 0.37 &16.24 & 0.03 & 15.41 & 0.04 & 14.85 & 0.04&  14.37&  0.05& K60\\ 
5831.20& 16.47 & 0.18 &16.27 & 0.04 & 15.41 & 0.04 & 14.91 & 0.03&  14.39&  0.04& K60\\ 
5832.20& 16.88 & 0.12 &16.32 & 0.04 & 15.46 & 0.04 & 14.90 & 0.04&  14.45&  0.04& K60\\ 
5835.19&       &      &16.50 & 0.05 & 15.52 & 0.03 & 14.90 & 0.02&       &      & K100a\\ 
5853.21&       &      &      &      & 15.84 & 0.12 & 15.29 & 0.10&  14.62&  0.20& S15\\ 
5853.22&       &      &      &      & 15.95 & 0.06 & 15.31 & 0.04&       &      & S60\\ 
5854.22&       &      &      &      & 15.92 & 0.06 & 15.48 & 0.04&  14.97&  0.12& S60\\ 
5856.20&       &      &      &      & 15.82 & 0.13 & 15.43 & 0.09&  14.87&  0.07& S15\\ 
5856.21&       &      &16.77 & 0.06 & 15.96 & 0.05 & 15.42 & 0.04&  14.97&  0.10& S60\\ 
5880.68&       &      &17.43 & 0.21 & 16.66 & 0.12 & 16.09 & 0.10&       &      & S15\\ 
5880.69&       &      &      &      & 16.56 & 0.08 &       &     &       &      & S60\\ 
5953.46&       &      &18.28 & 0.17 & 17.64 & 0.20 & 17.12 & 0.07&  16.99&  0.11& M70\\ 
5955.66&       &      &18.45 & 0.05 & 18.22 & 0.06 & 17.29 & 0.04&       &      & S60\\ 
5987.58&       &      &      &      &       &      & 17.91 & 0.07&       &      & S60\\ 
6022.39&       &      &      &      & 19.44 & 0.12 & 18.35 & 0.04&       &      & K100b\\ 
\end{longtable}
\end{center}                              

\bigskip
\bigskip
\PZsubtitle{\bf Light and color curves}
\PZfig{13cm}{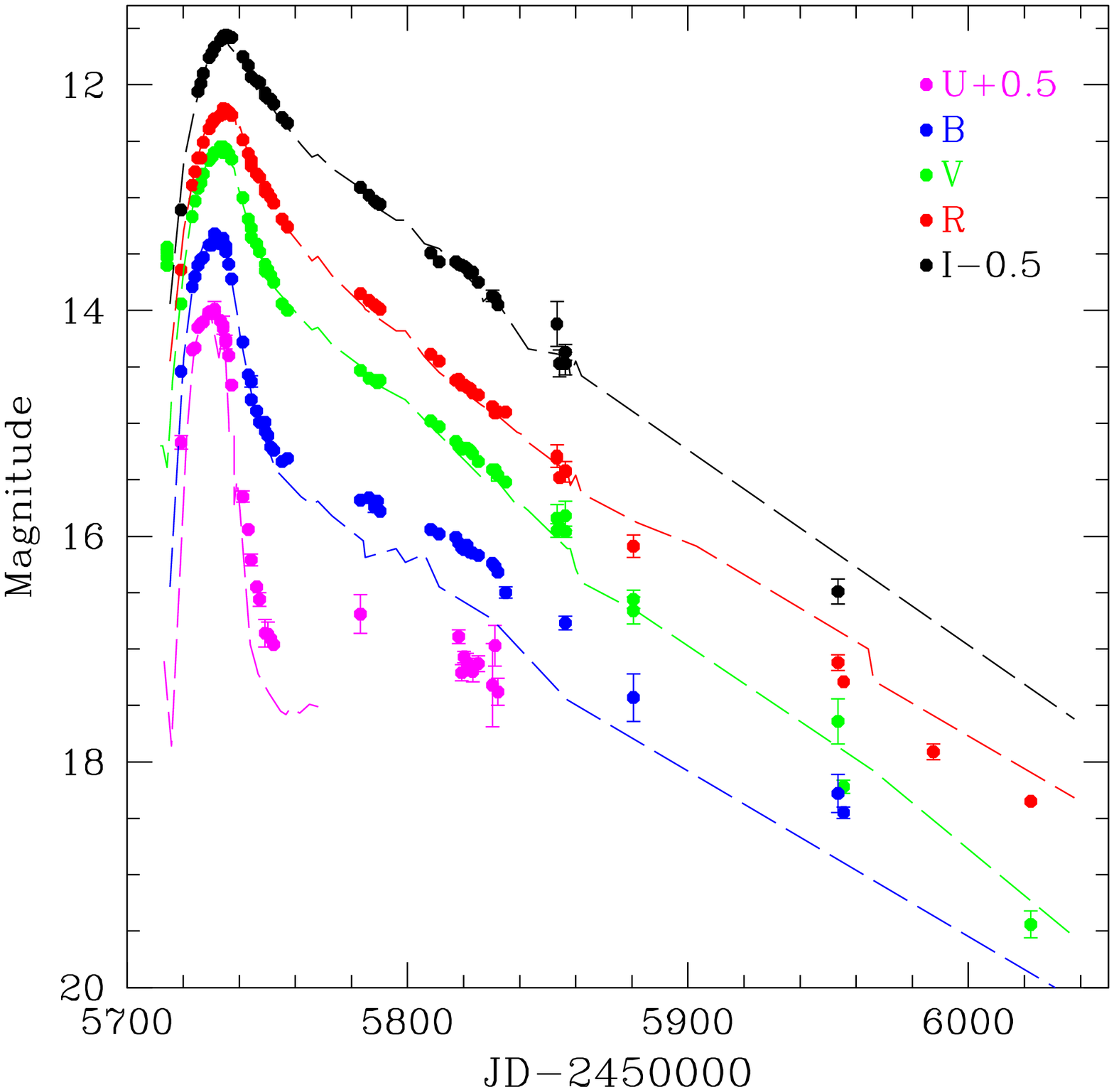}{The light curves of SN 2011dh. The dashed lines
are the light curves of SN 2008ax}
The light curves of SN 2011dh are shown in Fig. 2. The premaximum 
rise and the main peak have good coverage by observations, and we
can determine the dates and magnitudes of maximum light in different
bands: $U_{max}=13.49; t_{Umax}={\rm JD}2455730.4;
B_{max}=13.35; t_{Bmax}={\rm JD}2455732.3; V_{max}=12.56;
t_{Vmax}={\rm JD}2455733.9; R_{max}=12.22; t_{Rmax}={\rm JD}2455734.6;
I_{max}=12.06; t_{Imax}={\rm JD}2455735.6$.
After the maximum the brightness of SN declined very fast.
At the phase 15 days past maximum the $B$ magnitude declined by
1.64 mag. The fast drop continued for about 21 days, and at about 
JD 2455753 the onset of the linear decline is observed (K-point). 
The rates of decline in the period JD 2455780-2456010 are 
(in mag/day): 0.016 in $B$, 0.020 in $V$, 0.019 in $R$, 0.021 in $I$.

Comparison with SN 2008ax (Pastorello et al., 2008;
Tsvetkov et al., 2009) reveals good match of the light curves at the
main peak. After K-point the agreement is good in the $V$, $R$ and
$I$ bands, while in the $U$ and $B$ the luminosity decline of
SN 2011dh is slower. In the $U$ band there
is even a slight increase of brightness after K-point, and in 
the $B$ band a protrusion on the light curve can be noticed. 

\PZfig{13cm}{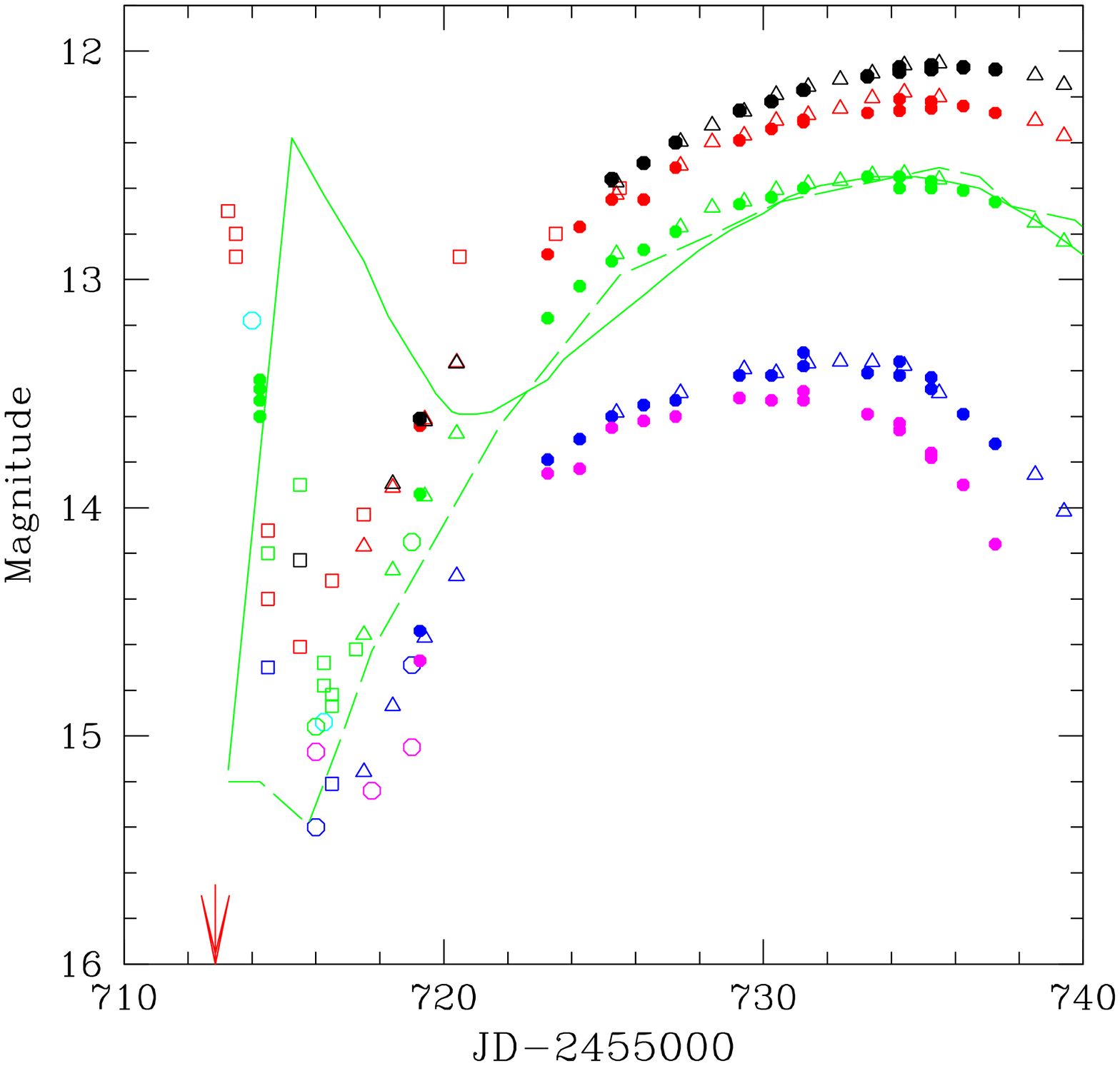}{The light curves of SN 2011dh for the first 30
days past outburst. The color coding is the same as in Fig.2,
but no magnitude shifts were applied. Observations in $g$ band are
plotted in cyan. We show our data (dots), and photometry by
Vinko et al. (2012) (triangles), Arcavi et al (2011a) (circles) and
by amateur astronomers (squares). Solid and dashed green lines are
the $V$-light curves of SNe 1993J and 2008ax}

Fig. 3 shows the light curves for the first 30 days after outburst.
We plotted our data and the observations by Vinko et al. (2011),
Arcavi et al. (2011a), and results of amateur astronomers,
taken from "Latest supernovae" site\PZfm   
\PZfoot{www.rochesterastronomy.org/supernova.html}

The last image of M51 with no SN visible (mag $>$ 18) was obtained
on JD 2455712.86, and the first detection was on JD 2455713.34.
We assume JD 2455712.9 as the time of explosion.
The initial flash was very fast: the rise to first peak
with brightness of about 12.8 mag 
took only 
about 0.4 days, and atfer 2.6 days the local minimum was reached
on JD 2455716, 
at $V$ about 15 mag.

We compare the early light curves of SN 2011dh with those for 
SNe IIb 1993J and 2008ax (Richmond et al., 1996; Pastorello et al.,
2008; Tsvetkov et al., 2009). The light curves were
shifted in time to coinside at the estimated moment of explosion,
and the shift in magnitudes was applied to match the main peak brightness.
The difference between the objects
is evident: the initial peak for SN 1993J was the widest and strongest
among these objects, while for SN 2008ax it was very weak.

\PZfig{13cm}{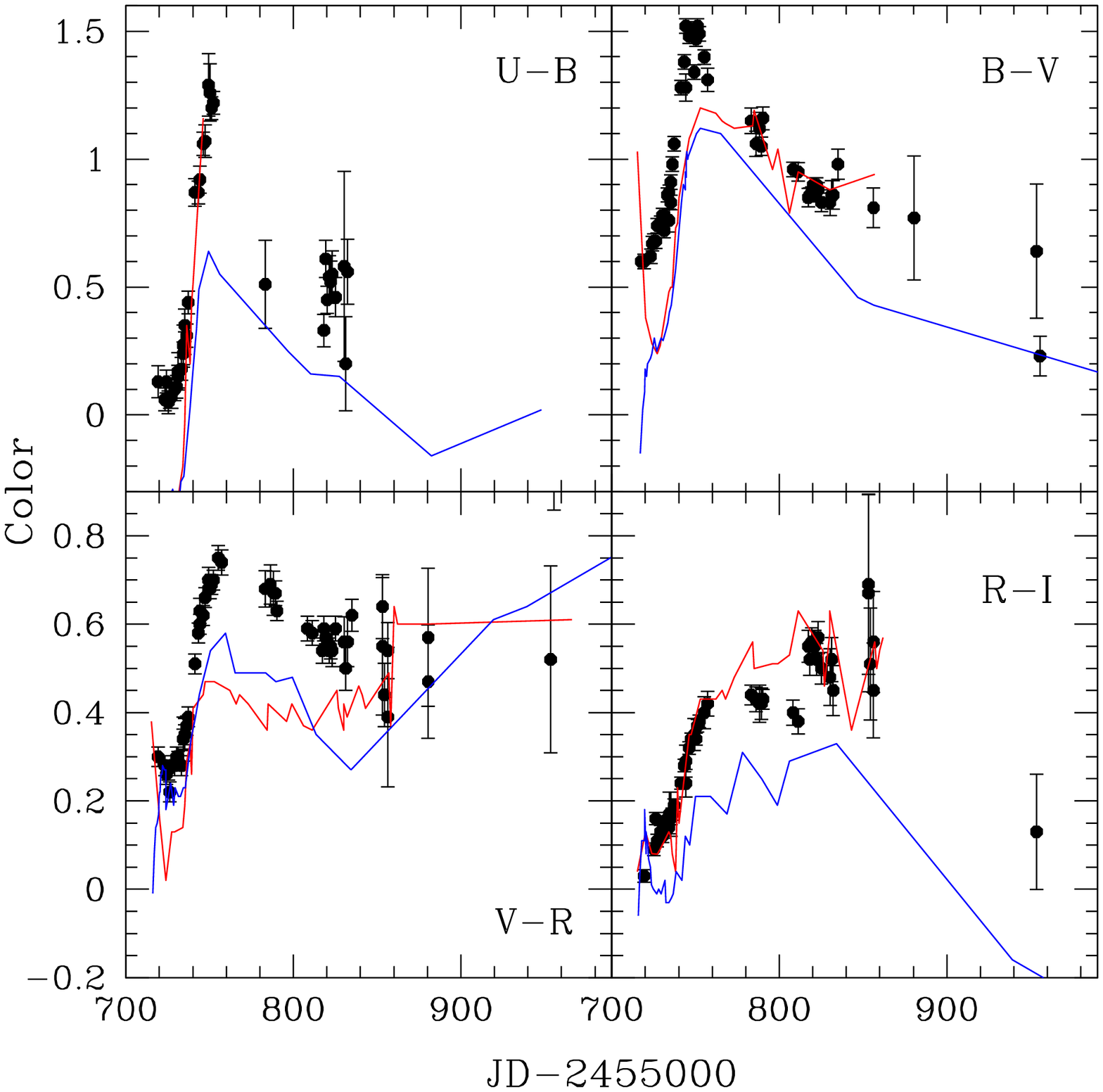}{The color curves of SN 2011dh. Blue and red lines are
the color curves of SNe 1993J and 2008ax}
 
The color curves are shown in Fig. 5. The evolution of colors
$U-B$, $B-V$ and $V-R$ is similar. SN 2011dh quickly reddens until K-point,
and then becomes bluer. The color $R-I$ remains nearly constant after 
K-point. 
The comparison with type IIb SNe 1993J and 2008ax reveals diversity
of the color curves, both in shape and the values of colors. Good match
is observed only for $U-B$ and $R-I$ colors between SNe 2011dh and
2008ax. SN 2011dh is significantly redder in $B-V$ and $V-R$ than the
other two objects.
 
The absolute $V$-magnitude light curves of SN 2008ax and several SNe
of types IIb, Ib and Ic are compared in Fig. 5.
\PZfig{13cm}{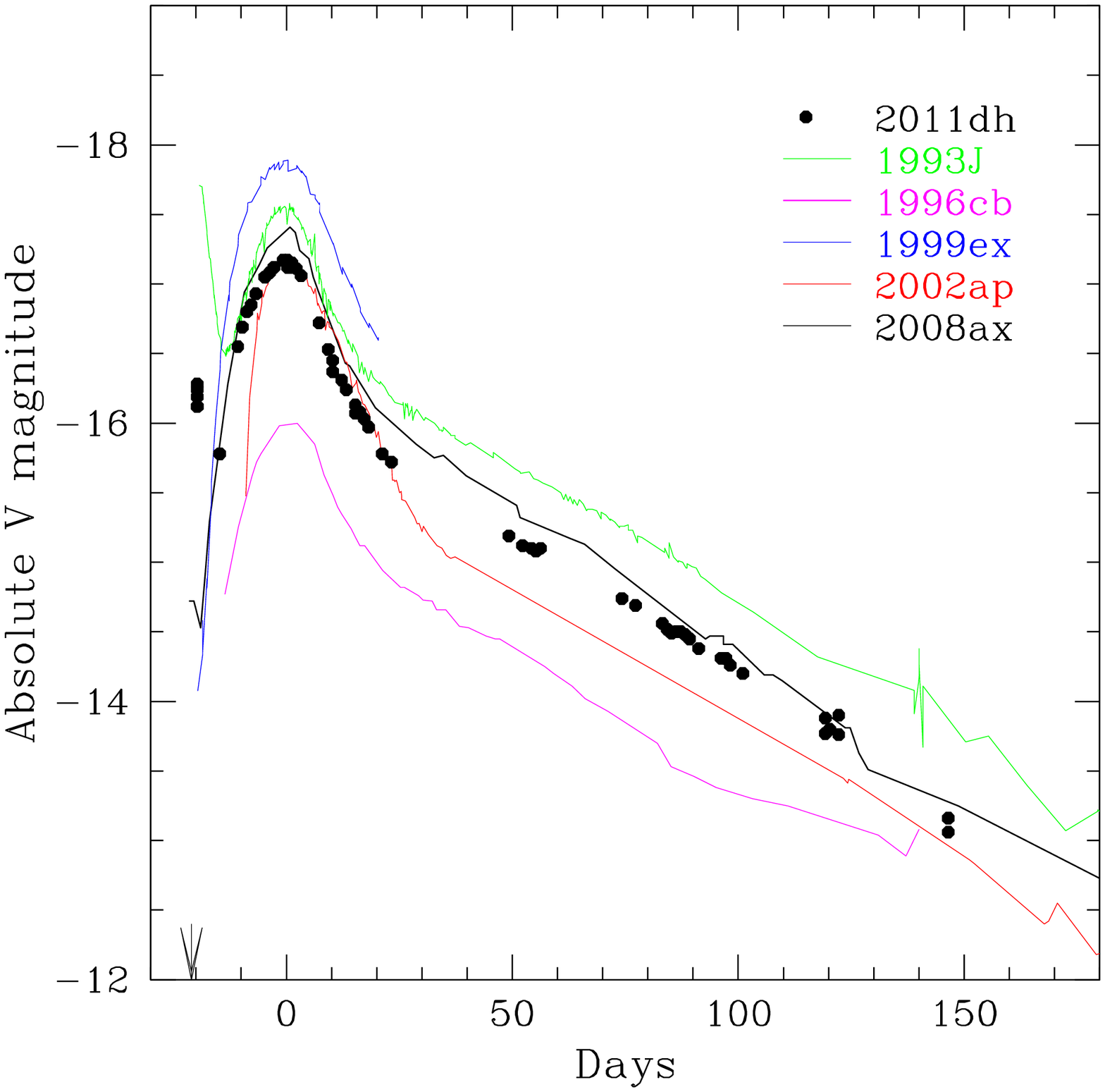}{The absolute $V$-band light
curves of SN 2011dh and SNe of types IIb (1993J, 1996cb, 2008ax), 
Ib (1999ex), Ic (2002ap). Day 0 corresponds to the main maximum on the
light curves} 
For SN 2011dh we adopted distance of 8.4 Mpc (Vinko et al., 2012) and
extinction $E(B-V)=0.035$. The light curves of other 
SNe are taken from Richmond et al. (1996), 
Qiu et al. (1999), Stritzinger et al. (2002), Foley et al. (2003),
Pastorello et al.
(2008), Tsvetkov et al. (2009).
With absolute peak $V$ magnitude of $-17.16$ mag SN 2011dh appears to 
be quite typical among SNe of similar classes. It is little fainter than
SNe IIb 1993J, 2008ax and SN Ib 1999ex, have nearly the same luminosity as 
SN Ic 2002ap and 
is significantly brighter than SN IIb 1996cb.

\bigskip
\bigskip
\PZsubtitle{\bf Modeling the light curves}
We derived quasi-bolometric light curve for SN 2011dh, integrating
the flux in {\it UBVRI} bands. On the dates when observations in some
bands were missing, we used the color curves to estimate the color of 
SN at that date and then calculated the missing magnitudes.
We attempted to model the quasi-bolometric light curve as well as
the light curves in
{\it UBVRI} bands using
our code STELLA, which incorporates implicit hydrodynamics coupled to
a time-dependent multi-group non-equilibrium radiative transfer
(Blinnikov et al., 1998). 
The specific model employed here was Model 13C of Woosley et al. (1994). 
This model was derived from a 13 M$_{\odot}$ main sequence star that
lost most of its hydrogen envelope to a nearby companion.
We present results for 6 variants of the model with different
values of radius, explosion energy and ejected mass, which
are reported in Table 2.
The mass of $^{56}$Ni was fixed at 0.07 M$_{\odot}$. 
The results are presented in Figs. 6-9.

The influence of changing main parameters on the shape of resulting
light curve can be seen in Figs. 6,7. The reduction of radius leads to 
shortening of the primary flash, but at the same time it becomes 
weaker. Model 5 with increased energy of explosion show the worst
agreement with observational data. Models 3,4 and 5 have good agreement
with observed curve at late stages.
Figs. 8,9 show the computed {\it UBVRI} light curves for the models 3 and 6,
which fit better the quasi-bolometric light curves.
The main maximum in the $V$ and $R$ bands is reproduced
satisfactorily, but the computed duration of the initial flash
is longer, and its lumonosity is lower than observed. The agreement
in other bands is worse. We may conclude that, although our models
reproduce main features of the observed light curves, the agreement
is not satisfactory.
We continue the search 
for models which will give better fits. The results and more detailed
discussion of the properties of the models and their impact on the possible 
evolution of the progenitor will be published in a subsequent paper.

\begin{table}
\caption{Model parameters}\vskip2mm
\begin{center}
\begin{tabular}{cccc}
\hline
Model & Radius, $R_{\odot}$ & Mass, $M_{\odot}$ & Energy, 10$^{51}$ erg/s \\
\hline
1 & 562 & 2.24 & 1.5 \\
2 & 300 & 2.24 & 1.5 \\
3 & 300 & 2.24 & 2.0 \\
4 & 150 & 2.24 & 2.0 \\
5 & 150 & 2.24 & 4.0 \\
6 & 300 & 4.24 & 2.0 \\
\hline
\end{tabular}
\end{center}
\end{table}

\newpage
\PZfig{13cm}{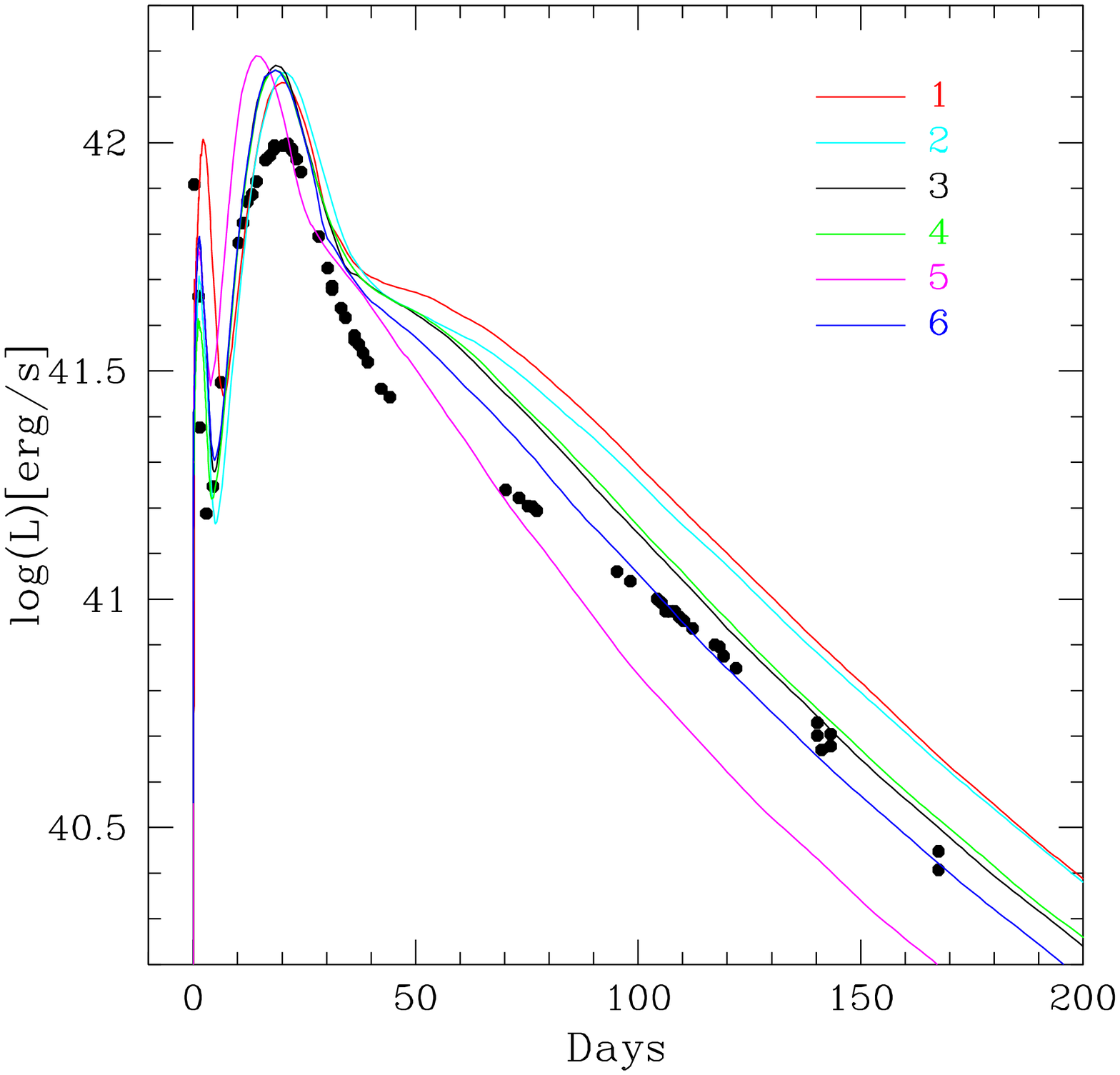}{The computed quasi-bolometric light curves for 6 
models compared to the quasi-bolometric light curve of SN 2011dh (black 
dots). Day 0 is JD 2455712.9}

\newpage
\PZfig{13cm}{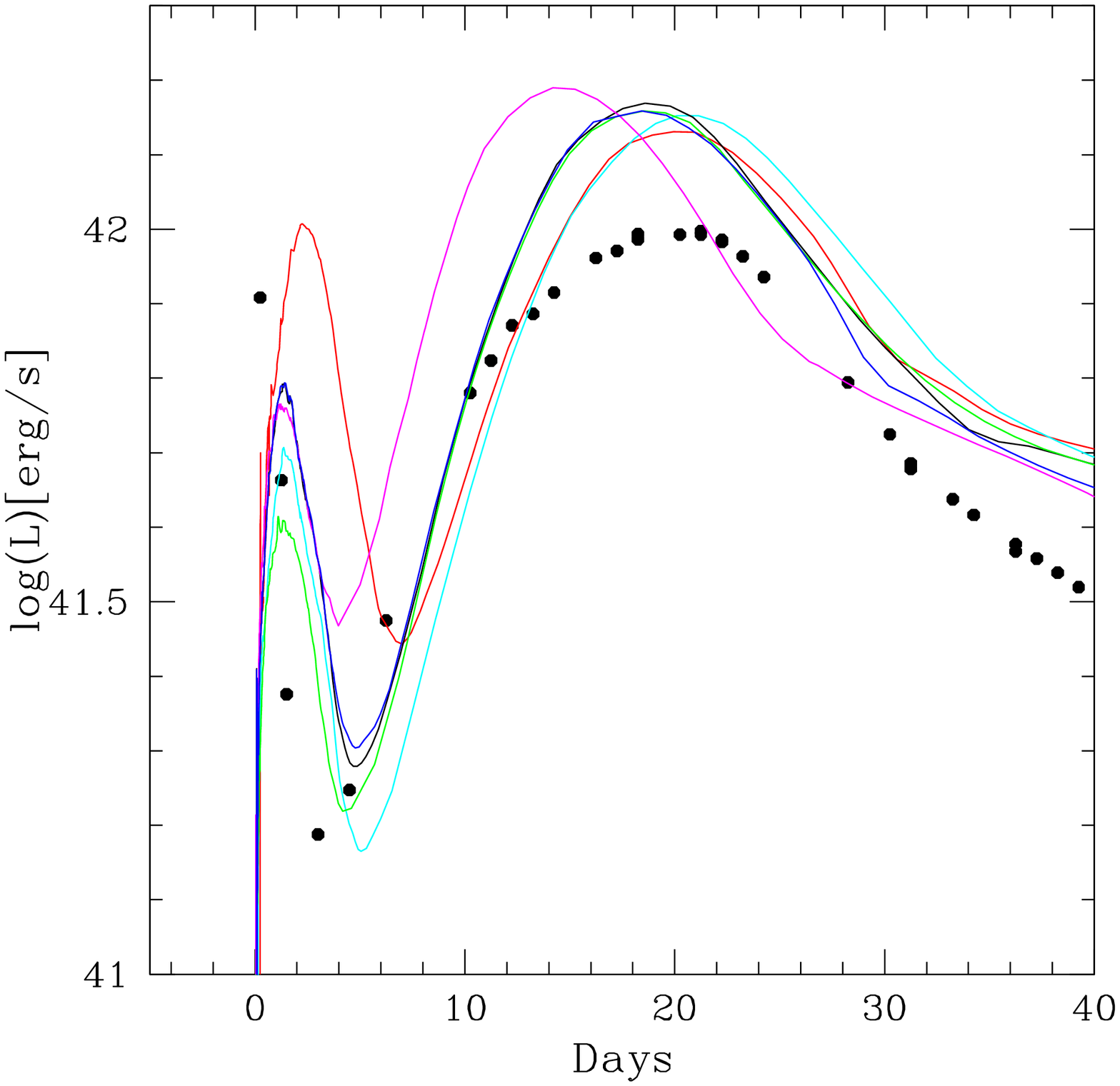}{The same as Fig. 6 for the first 40 days
past explosion}

\newpage
\PZfig{13cm}{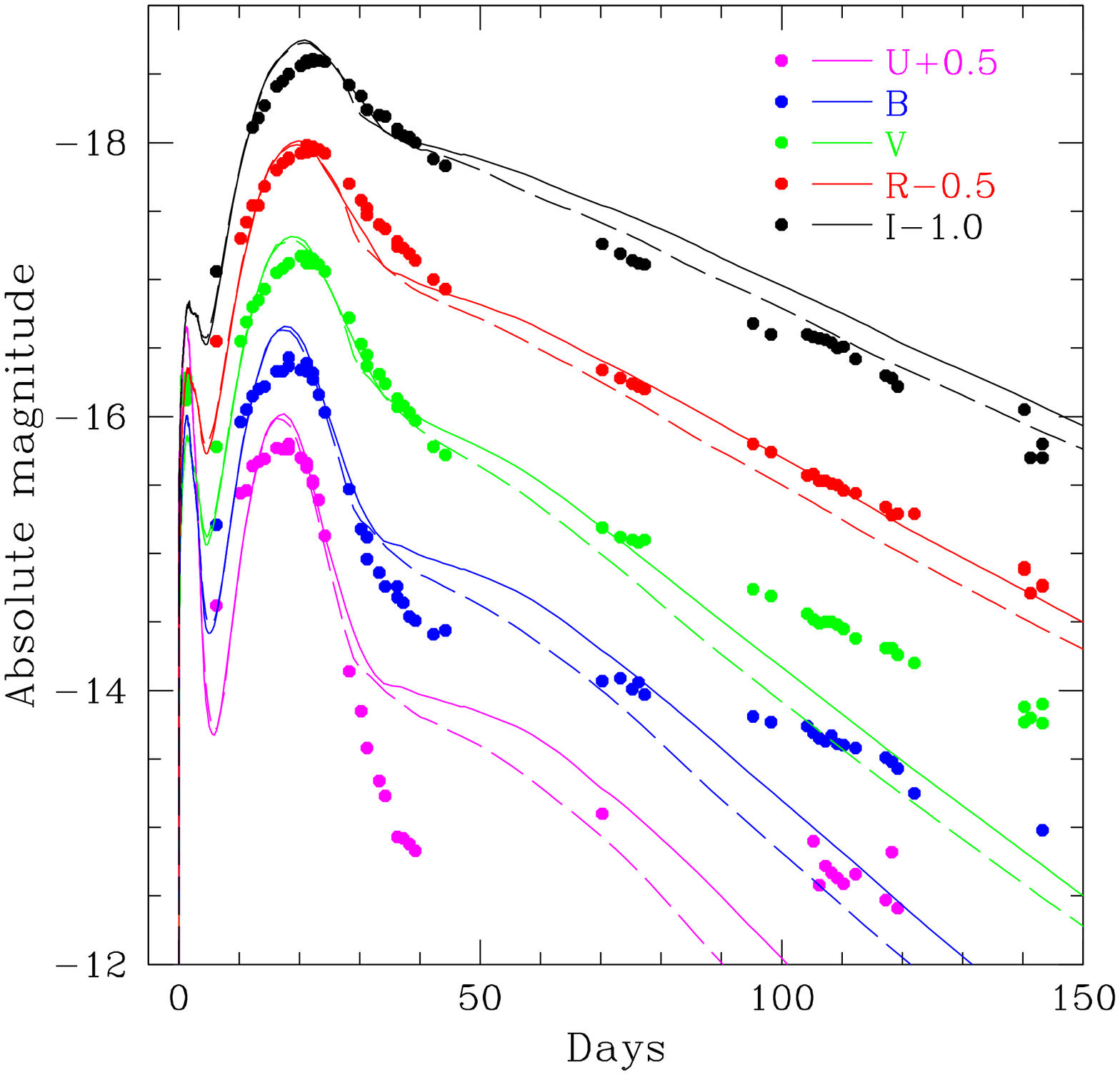}{The light curves for models 3 (solid lines)
and 6 (dashed lines) in {\it UBVRI} 
bands 
compared to the observed light curves of SN 2011dh}

\newpage
\PZfig{13cm}{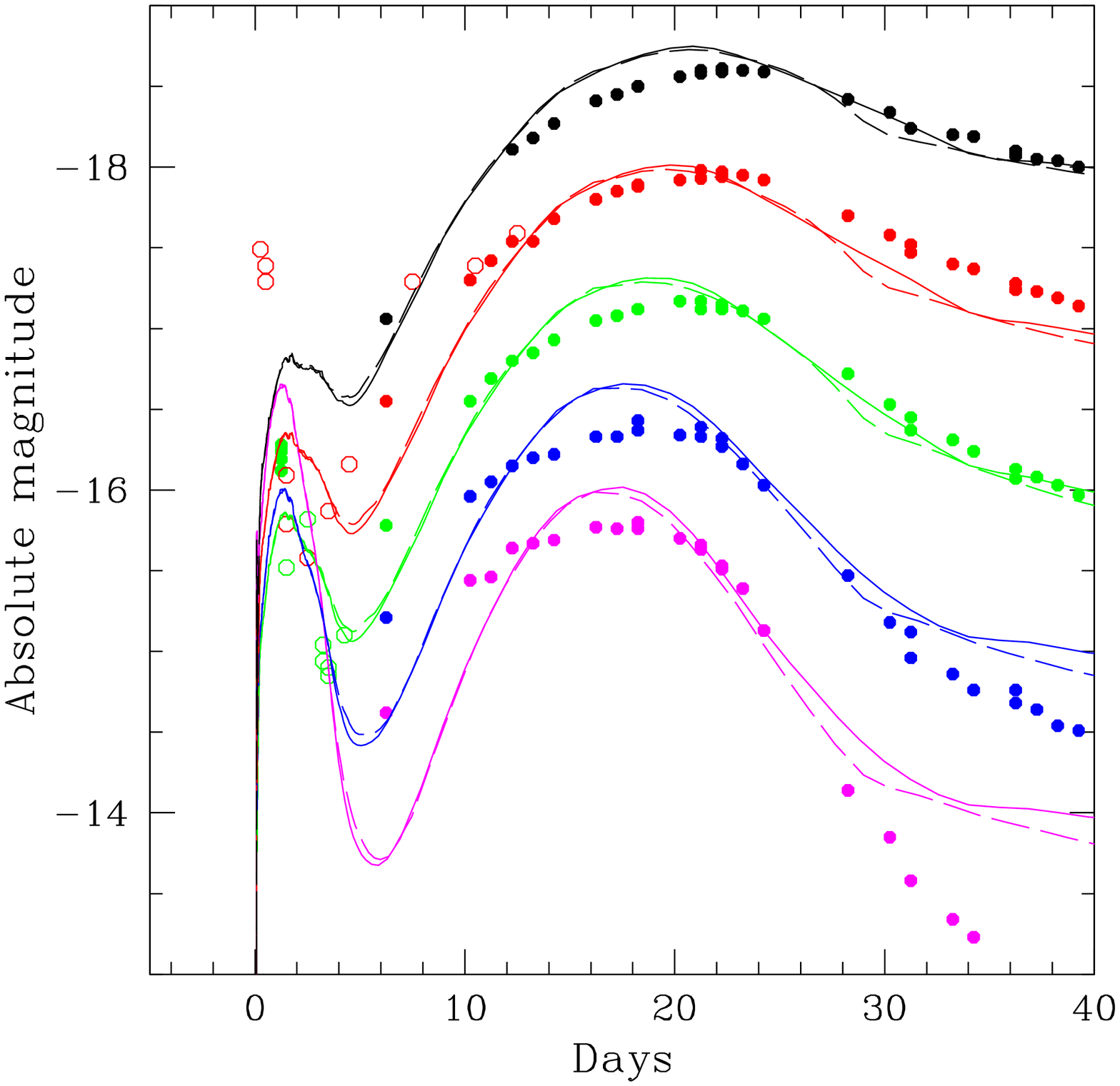}{The same as Fig. 8 for the first 40 days 
past explosion. Observatons of amateur astronomers are plotted as
circles}

\bigskip
{\bf Acknowledgements.} We thank N.P.Ikonnikova who
made some of the observations.

The work is supported partly by the grant 
of the Government of the Russian Federation (No 11.G34.31.0047),
by RFBR grants 10-02-00249a, 10-02-01398a, 11-02-01213a,
by RF Sci. Schools 3458.2010.2 and 3899.2010.2,
by the grant IZ73Z0-128180/1 of the Swiss National Science
Foundation (SCOPES), and by SAIA -- scholarship (Slovakia).

\references

Arcavi, I., Gal-Yam, A., Polishook, E., et al., 2011a,
{\it Astronomer's Telegram}, No.3413

Arcavi, I., Gal-Yam, A., Yaron, O.,  et al., 
2011b, {\it Astrophys. J.}, 742, L18 

Bietenholz, M.F., Brunthaler, A., Soderberg, A.M., et al., 2012, 
{\it Astrophys. J.}, 751, 125 

Blinnikov, S. I., Eastman, R., Bartunov, O. S., 
Popolitov, V. A., Woosley, S. E., 1998, 
{\it Astrophys. J.}, 496, 454 

Foley, R.J., Papenkova, M.S., Swift, B.J., et al.,
2003, {\it PASP}, 115, 1220

Krauss, M.I., Soderberg, A.M., Chomiuk, L., et al., 2012,
{\it Astrophys. J.}, 750, L40  

Li, W., Filippenko, A.V., 2011, {\it Astronomer's Telegram}, No.3399

Marti-Vidal, I., Tudose, V., Paragi, Z., et al., 2011,
{\it Astron. Astrophys.}, 535, L10  

Maund, J.R., Fraser, M., Ergon, M., et al., 2011,
{\it Astrophys. J.}, 739, L37  

Pastorello, A., Kasliwal, M.M., Crockett, R.M., et al., 2008,
{\it MNRAS}, 389, 955

Pastorello, A., Valenti, S., Zampieri, L., et al., 2009, {\it MNRAS},
394, 2266

Richmond, M.W., Treffers, R.R., Filippenko, A.V., Paik, Y., 1996,
{\it Astron. J.}, 112, 732

Soderberg, A.M., Margutti, R., Zauderer, B.A., et al., 2012, 
{\it Astrophys. J.}, 752, 78  

Stritzinger, M,, Hamuy, M., Suntzeff, N.B., et al.,
2002, {\it Astron. J.}, 124, 2100

Szczygiel, D.M., Gerke, J.R., Kochanek, C.S., Stanek, K.Z., 2012,
{\it Astrophys. J.}, 747, 23  

Tsvetkov, D.Yu., Volkov, I.M., Baklanov, P.V., Blinnikov, S.I.,
Tuchin, O., 2009, {\it Variable Stars}, 29, No.2

Van Dyk, S.D., Li, W., Cenko, S.B., et al., 2011,
{\it Astrophys. J.}, 741, L28  

Vinko, J., Takats, K., Szalai, T., et al., 2012, 
{\it Astron. Astrophys.}, 540, 93  

Woosley, S.E., Eastman, R.G., Weaver, T.A., Pinto, P.A., 1994,
{\it Astrophys. J.}, 429, 300 

Qiu, Y., Li, W., Qiao, Q., Hu, J., 1999, {\it Astron. J.}, 117, 736
 
\endreferences

\end{document}